\documentclass[aps,prd,nofootinbib,showpacs]{revtex4}
\usepackage{amssymb}
\usepackage{amsmath}
\usepackage{amsfonts}
\usepackage{epsfig}
\usepackage{appendix}
\def\uu{\langle \bar u u \rangle}
\def\dd{\langle \bar d d \rangle}

\newcommand{\seq}{\begin{subequations}}
\newcommand{\sen}{\end{subequations}}
\newcommand{\eq}{\begin{eqnarray}}
\newcommand{\en}{\end{eqnarray}}

\begin{document}

\title{Tree Level  Semileptonic $\Sigma_{b}$ to Nucleon Decay in Light Cone QCD Sum Rules }
\author{K. Azizi$^{1}$\footnote{e146342@metu.edu.tr},
        M. Bayar$^{2}$\footnote{melahat.bayar@kocaeli.edu.tr},
       A. Ozpineci$^{1}$\footnote{ozpineci@metu.edu.tr},
        Y. Sarac$^3$\footnote{ysoymak@atilim.edu.tr},
      }

\affiliation{\vspace*{1.2\baselineskip}\\$^1$ Physics
Department, Middle East Technical University, 06531, Ankara, Turkey\\
\vskip 1pt
$^2$ Department of Physics, Kocaeli University, 41380 Izmit, Turkey\\
\vskip 1pt
$^3$ Electrical and Electronics Engineering Department, Atilim University, 06836 Ankara, Turkey\\
}
\begin{titlepage}
\date{\today}

\begin{abstract}
 Using the most general form of the interpolating current of the heavy spin 1/2, $\Sigma_{b}$ baryon and distribution amplitudes
 of the nucleon, the transition form factors of the semileptonic $\Sigma_{b}\rightarrow Nl\nu$ decay are calculated in the framework of  light
 cone QCD sum rules. It is obtained that the form factors satisfy the heavy quark effective theory relations. The obtained results for the related form factors are
 used to estimate the decay rate of this transition.
\end{abstract}


\maketitle

\end{titlepage}

\section{Introduction}
The baryons containing a heavy quark have been at the focus of much
theoretical attention, especially  since the development of the
heavy quark effective theory (HQET) and its application to the
spectroscopy of these baryons. The heavy quark provides a window
that permits  us to see further under the skin of the
non-perturbative QCD as compared the light baryons. These states are
expected to be narrow, so that their isolation and detection are
relatively easy. Recently, experimental studies on the spectroscopy
of these baryons have been accelerated and  new  heavy baryons have
been discovered
\cite{Mattson,Ocherashvili,Acosta,Chistov,Aubert1,Abazov1,Aaltonen1,Solovieva}.
The $\Sigma_b$ channels are expected to be very rich, so it will be
possible to check its semileptoic decays like its decay to the
nucleon  at LHC in the  near future. There are many works in
literature which are devoted to the investigation of the mass and
magnetic moments of the heavy baryons using different approaches.
The masses of these baryons have been discussed within QCD sum rules
in \cite{kazem1,Shuryak,Kiselev1,
Kiselev2,Bagan1,Bagan3,Duraes,Wang1,Zhang1}, in heavy quark
effective theory (HQET) in
\cite{Grozin,Groote,Dai,Lee,Huang,Liu,Wang2} and using different
quark models in
\cite{Ebert1,Ebert3,Capstick,Matrasulov,Gershtein,Kiselev3,Vijande,Martynenko,Hasenfratz}.
The magnetic dipole moment of heavy spin 1/2 and 3/2 baryons as well
as the transition magnetic dipole and electric quadrupole moments of
heavy spin 3/2 to heavy spin 1/2 baryons have been calculated in the
framework of different approaches ( see for example
\cite{kazem1,kazem2,kazem3} and references therein). However, the
semileptonic and nonleptonic decays of the heavy baryons have not
been extensively discussed in the literature comparing their  mass and
electromagnetic properties. Transition form factors of the   $\Lambda_b\rightarrow \Lambda_c$ and  $\Lambda_c\rightarrow \Lambda$  decays have been studied in three points QCD sum rules  in \cite{yeni1}, and then used in the study of the semileptonic decays. The  $\Lambda_{b}\rightarrow pl\bar\nu$ transition has also been investigated using three point QCD sum rules  within the framework of heavy quark effective theory (HQET) in \cite{yeni2} and using SU(3) symmetry and HQET in \cite{Datta}.  Hyperfine mixing and
the semileptonic decays of double-heavy baryons in a quark model
\cite{Roberts}, strong decays of heavy baryons in Bethe-Salpeter
formalism \cite{Guo}, strong decays of charmed baryons in heavy
hadron Chiral perturbation theory \cite{Cheng} and semileptonic decays of some heavy  baryons containing single heavy quark in different  quark models
\cite{D.Ebert,albertus,pervin} are some other works related to the heavy baryon decays.

In the present work, we calculate the form factors related to the
semileptonic decay of the $\Sigma_b\rightarrow   Nl\nu$ transition
in the framework of the light cone QCD sum rules using the nucleon
distribution amplitudes. Here, N refers to two members of the octet
baryons, namely neutron and proton. The parameters appearing in the
nucleon distribution amplitudes have been calculated using various
methods. In this work, for the values of these parameters, we use
the results of QCD sum rules approach \cite{Lenz} and  also the
results which are recently obtained from lattice QCD
\cite{Gockeler1,Gockeler2,QCDSF}. Analyzing of such transitions can
give essential information about the internal structure of the
$\Sigma_b$ baryon as well as accurate calculation of the nucleon
wave functions.  Since the spin of the heavy baryon carries
information on the spin of the heavy quark, the study of such
transitions might also lead us to study the spin effects in the
heavy quark sector of the standard model.

The outline of the paper is as follows: in section II,  using the
nucleon distribution amplitudes and the most general form of the
interpolating currents for the $\Sigma_b$ baryon,  we calculate the
form factors entering to the semileptonic decay of the heavy
$\Sigma_{b}$ baryon to nucleon  in the framework of the light cone
QCD sum rules. The heavy quark limit of the form factors and the
relations between the form factors in this limit is also discussed
in this section. Section III encompasses  numerical analysis of the
form factors, our predictions for the decay rate obtained in two
different ways: first, using the DA's obtained from QCD sum rules
and second, the DA's calculated in lattice QCD , and discussion.

\section{ Light cone QCD sum rules for the $\Sigma_{b}\rightarrow N$ form factors }
This section is devoted to  the calculation of  form factors
relevant for the $\Sigma_{b}^0 \rightarrow p$ and $\Sigma_{b}^-
\rightarrow n$ transitions using the light cone QCD sum rules
approach. At quark level, these transitions are governed by the tree
level $b\rightarrow u$  transition. Considering the SU(2) symmetry,
the form factors of these two transitions are the same, so we will
use the notation N instead of neutron and proton. The quark level
transition is described by the effective Hamiltonian given by
\begin{eqnarray}
{\cal H}_{eff} = \frac{G_F}{\sqrt2} V_{ub} \bar u \gamma_\mu
(1-\gamma_5) b \bar l \gamma^\mu (1-\gamma_5) \nu.
\end{eqnarray}
Hence, to study $\Sigma_b \rightarrow N l \nu$ decay, one needs the
matrix element $\langle N \vert \bar u \gamma_\mu (1-\gamma_5) b
\vert \Sigma_b \rangle $. To calculate this matrix element,
following the general philosophy of QCD sum rules, we start by
considering the  correlation function,
\begin{equation}\label{T}
\Pi_{\mu}(p,q)=i\int d^{4}xe^{iqx}\langle N(p)\mid
T\{J^{tr}_{\mu}(x)\bar J^{\Sigma_b}(0) \}\mid 0\rangle,
\end{equation}
where, $J^{\Sigma_b}$ is interpolating currents of $\Sigma_b$
baryon,  $J^{tr}_{\mu}=\bar u\gamma_{\mu}(1-\gamma_5)b$ is
transition current and $\langle N(p)\mid$ presents the proton sate.
$p$ denotes the proton momentum and $q=(p+q)-p$ is the transferred
momentum. To calculate the form factors, the following three steps
will be applied:
\begin{itemize}
\item The correlation function is calculated by saturating it with a
tower of hadrons having the same quantum number as the interpolating
current, $J^{\Sigma_b}$ called the phenomenological or physical
side.
\item The correlation function is calculated in QCD or theoretical
side via operator product expansion (OPE), where the short and long
distance quark-gluon interactions are separated. The former is
calculated using QCD perturbation theory, whereas the latter are
parameterized in terms of the light-cone distribution amplitudes of
the nucleon.
\item The sum rules for form factors are calculated equating the two representation of the correlation function
mentioned above and applying Borel transformation to suppress the
contribution of the higher states and continuum.

\end{itemize}

To calculate  the physical  side,   a complete set of hadronic state
is inserted to the correlation function. After performing integral
over x, we obtain
\begin{equation}\label{phys1}
\Pi_{\mu}(p,q)=\sum_{s}\frac{\langle N(p)\mid J^{tr}_{\mu}(x)\mid
\Sigma_b(p+q,s)\rangle\langle \Sigma_b(p+q,s)\mid
\bar J^{\Sigma_b}(0)\mid 0\rangle}{m_{\Sigma_b}^{2}-(p+q)^{2}}+...,
\end{equation}
where, the ... represents the contribution of the higher states and
continuum. The matrix element $\langle\Sigma_b(p+q,s)\mid
\bar J^{\Sigma_b}(0)\mid 0\rangle$ in (\ref{phys1}) can be written as:
\begin{equation}\label{matrixel2}
\langle\Sigma_b(p+q,s)\mid \bar J^{\Sigma_b}(0)\mid
0\rangle=\lambda_{\Sigma_b} \bar u_{\Sigma_b}(p+q,s),
\end{equation}
where $\lambda_{\Sigma_b}$ is residue of $\Sigma_b$ baryon. The
transition  matrix element, $\langle N(p)\mid J_{\mu}^{tr}\mid
\Sigma_b(p+q,s)\rangle$ is parameterized in terms of the form
factors $f_{i}$ and $g_{i}$ as

\begin{eqnarray}\label{matrixel1}
\langle N(p)\mid J_{\mu}^{tr}(x)\mid \Sigma_b(p+q)\rangle&=&\bar
N(p)\left[\gamma_{\mu}f_{1}(Q^{2})+{i}\sigma_{\mu\nu}q^{\nu}f_{2}(Q^{2})+
q^{\mu}f_{3}(Q^{2})+\gamma_{\mu}\gamma_5
g_{1}(Q^{2})+{i}\sigma_{\mu\nu}\gamma_5q^{\nu}g_{2}(Q^{2})\right.\nonumber\\
&+& \left. q^{\mu}\gamma_5 g_{3}(Q^{2})
\vphantom{\int_0^{x_2}}\right] u_{\Sigma_b}(p+q),\nonumber\\
\end{eqnarray}
where $Q^{2}=-q^{2}$, and $f_{i}$, and $g_{i}$,  are the form
factors and $N(p)$ and $u_{\Sigma_b}(p+q)$ are the spinors of
nucleon and $\Sigma_b$, respectively.  Using Eqs. (\ref{phys1}),
(\ref{matrixel2}) and ,(\ref{matrixel1}) and summing over spins of
the $\Sigma_b$ baryon using
\begin{equation}\label{spinor}
\sum_{s}u_{\Sigma_b}(p+q,s)\overline{u}_{\Sigma_b}(p+q,s)=\not\!p+\not\!q+m_{\Sigma_b},
\end{equation} we obtain the following expression
\begin{eqnarray}\label{phys2}
\Pi_{\mu}(p,q)&=&
\frac{\lambda_{\Sigma_b}}{m_{\Sigma_b}^{2}-(p+q)^{2}}\bar
N(p)\left[\gamma_{\mu}f_{1}(Q^{2})+{i}\sigma_{\mu\nu}q^{\nu}f_{2}(Q^{2}+
q^{\mu}f_{3}(Q^{2})+\gamma_{\mu}\gamma_5
g_{1}(Q^{2})+{i}\sigma_{\mu\nu}\gamma_5q^{\nu}g_{2}(Q^{2})\right.\nonumber\\
&+& \left. q^{\mu}\gamma_5 g_{3}(Q^{2})
\vphantom{\int_0^{x_2}}\right] (\not\!p+\not\!q+m_{\Sigma_b}) +
\cdots
\end{eqnarray}
 Using
\begin{eqnarray}\label{sigma}
\bar{N}\sigma_{\mu\nu}q^{\nu}u_{\Sigma_b}&=&
\bar{N}[(m_N+m_{\Sigma_b})\gamma_{\mu}-(2p+q)_\mu]u_{\Sigma_b},
\end{eqnarray}
in Eq. (\ref{phys2}), the final expression  for the physical side of
the correlation function is obtained as
\begin{eqnarray}\label{sigmaafter}
\Pi_{\lambda}(p,q)&=&
\frac{\lambda_{\Sigma_b}}{m_{\Sigma_b}^{2}-(p+q)^{2}}\bar
N(p)\left[\vphantom{\int_0^{x_2}}2f_{1}(Q^{2})p_\mu+\left\{\vphantom{\int_0^{x_2}}-f_1(Q^2)(m_N-m_{\Sigma_b})
+f_2(Q^2)(m_N^2-m_\Sigma^2)\right\}\gamma_\mu\right.\nonumber \\
&&+\left\{\vphantom{\int_0^{x_2}}f_1(Q^2)-f_2(Q^2)(m_N+m_{\Sigma_b})\right\}\gamma_\mu\not\!q+
2f_{2}(Q^{2})p_\mu\not\!q
+\left\{\vphantom{\int_0^{x_2}}f_2(Q^2)+f_3(Q^2)\right\}(m_N+m_{\Sigma_b})q_\mu\nonumber\\&+&\left\{\vphantom{\int_0^{x_2}}f_2(Q^2)
+f_3(Q^2)\right\}q_\mu\not\!q- 2g_1(Q^2)p_{\mu}\gamma_5
+\left\{\vphantom{\int_0^{x_2}}g_1(Q^2)(m_N+m_{\Sigma_b})\right.-\left.g_2(Q^2)(m_N^2-m^2_{\Sigma_b})\vphantom{\int_0^{x_2}}\right\}
\gamma_\mu\gamma_5-\nonumber\\&&
\left\{\vphantom{\int_0^{x_2}}g_{1}(Q^{2})-g_2(Q^2)(m_N-m_{\Sigma_b})\right\}\gamma_\mu\not\!q\gamma_5-2g_2(Q^2)p_\mu\not\!q\gamma_5
-\left\{\vphantom{\int_0^{x_2}}g_2(Q^2)+g_3(Q^2)\right\}(m_N-m_{\Sigma_b})q_\mu\gamma_5\nonumber\\&&-\left\{g_2(Q^2)+g_3(Q^2)\vphantom{\int_0^{x_2}}\right\}q_\mu\not\!q\gamma_5\left.\vphantom{\int_0^{x_2}}\right]+
\cdots
\end{eqnarray}
Among many structures appearing in Eq. (\ref{phys2}),  we chose the
independent structures  $p_{\mu}$, $p_{\mu}\!\!\not\!q$,
$q_{\mu}\!\!\not\!q$, $p_{\mu}\gamma_5$,
$p_{\mu}\!\!\not\!q\gamma_5$, and $q_{\mu}\!\!\not\!q\gamma_5$  to
evaluate the form factors $f_{1}$, $f_{2}$, $f_{3}$,  $g_{1}$,
$g_{2}$ and  $g_{3}$, respectively.

On     QCD     side, to calculate the correlation function in deep
Euclidean region where $(p+q)^2\ll0$, we need to know the explicit
expression for the interpolating current of the $\Sigma_b$ baryon.
It is chosen as

\begin{eqnarray}\label{cur.N}
J^{\Sigma_b}(x)&=&
\frac{-1}{\sqrt{2}}\varepsilon^{abc}\left[\vphantom{\int_0^{x_2}}\left\{\vphantom{\int_0^{x_2}}u^{T
a} (x)C b^{b} (x) \right\}\gamma_5 d^{c}
(x)-\left\{\vphantom{\int_0^{x_2}}b^{T a} (x)C d^{b} (x)
\right\}\gamma_5
u^{c} (x)\right. \nonumber \\
&& \left. +\beta\left\{\vphantom{\int_0^{x_2}}\{u^{T a} (x)C
\gamma_5 b^{b} (x) \} d^{c} (x)-\{b^{T a} (x)C \gamma_5 d^{b} (x) \}
u^{c} (x)\right\}\vphantom{\int_0^{x_2}}\right],
\end{eqnarray}
where $a,~b,~c$ are the color indices and $C$ is the charge conjugation
operator and $\beta$ is an arbitrary parameter with $\beta=-1$
corresponding to the Ioffe current. Using the transition current,
$J^{tr}_{\mu}=\bar u\gamma_{\mu}(1-\gamma_5)b$ and $J^{\Sigma_b}
$and contracting out all quark pairs applying the Wick's theorem, we
obtain
\begin{eqnarray}\label{mut.m}
\Pi_\mu &=& \frac{-i}{\sqrt{2}} \epsilon^{abc}\int d^4x e^{iqx}
\Bigg\{\Big[( C )_{\eta\lambda} (\gamma_5)_{\gamma\phi}-( C
)_{\lambda\phi} (\gamma_5)_{\gamma\eta}\Big]  +\beta\Bigg[(C
\gamma_5 )_{\eta\lambda}(I)_{\gamma\phi}
 \nonumber \\
&-& (C \gamma_5 )_{\lambda\phi}(I)_{\gamma\eta} \Bigg]\Bigg\} \Big[
(1+\gamma_5)\gamma_{\mu} \Big]_{\sigma\theta}S_Q(-x)_{\lambda\sigma}
\langle  N (p) | \bar u_\eta^a(0)
\bar u_\theta^b(x)  \bar d_\phi^c(0) | 0\rangle ,\nonumber\\
\end{eqnarray}
where, $ S_Q(x)$ is the heavy quark propagator which is represented
as \cite{Balitsky}:

\begin{eqnarray}\label{heavylightguy}
 S_Q (x)& =&  S_Q^{free} (x) - i g_s \int \frac{d^4 k}{(2\pi)^4}
e^{-ikx} \int_0^1 dv \Bigg[\frac{\not\!k + m_Q}{( m_Q^2-k^2)^2}
G^{\mu\nu}(vx) \sigma_{\mu\nu} + \frac{1}{m_Q^2-k^2} v x_\mu
G^{\mu\nu} \gamma_\nu \Bigg].
 \end{eqnarray}
where
\begin{eqnarray}\label{freeprop}
S^{free}_{Q}
&=&\frac{m_{Q}^{2}}{4\pi^{2}}\frac{K_{1}(m_{Q}\sqrt{-x^2})}{\sqrt{-x^2}}-i
\frac{m_{Q}^{2}\not\!x}{4\pi^{2}x^2}K_{2}(m_{Q}\sqrt{-x^2}),\nonumber\\
\end{eqnarray}
and  $K_i$ are the Bessel functions. The terms proportional to the
gluon strength tensor can give contribution to four and five
particle distribution functions but they are expected to be small
\cite{17,18,Braun1b} and for this reason, we will neglect these
amplitudes in further analysis.

For the calculation of $\Pi_\mu$ in Eq. (\ref{mut.m}), the matrix
element $\langle  N (p)\mid \epsilon^{abc}\bar u_{\eta}^{a}(0)\bar
u_{\theta}^{b}(x)\bar d_{\phi}^{c}(0)\mid 0\rangle$ is required. The
nucleon wave function is given as \cite{Lenz,17,18,Braun1b,8}:
\begin{eqnarray}\label{wave func}
&&4\langle0|\epsilon^{abc}u_\alpha^a(a_1 x)u_\beta^b(a_2
x)d_\gamma^c(a_3 x)|N(p)\rangle\nonumber\\
&=&\mathcal{S}_1m_{N}C_{\alpha\beta}(\gamma_5N)_{\gamma}+
\mathcal{S}_2m_{N}^2C_{\alpha\beta}(\rlap/x\gamma_5N)_{\gamma}\nonumber\\
&+& \mathcal{P}_1m_{N}(\gamma_5C)_{\alpha\beta}N_{\gamma}+
\mathcal{P}_2m_{N}^2(\gamma_5C)_{\alpha\beta}(\rlap/xN)_{\gamma}+
(\mathcal{V}_1+\frac{x^2m_{N}^2}{4}\mathcal{V}_1^M)(\rlap/pC)_{\alpha\beta}(\gamma_5N)_{\gamma}
\nonumber\\&+&
\mathcal{V}_2m_{N}(\rlap/pC)_{\alpha\beta}(\rlap/x\gamma_5N)_{\gamma}+
\mathcal{V}_3m_{N}(\gamma_\mu
C)_{\alpha\beta}(\gamma^\mu\gamma_5N)_{\gamma}+
\mathcal{V}_4m_{N}^2(\rlap/xC)_{\alpha\beta}(\gamma_5N)_{\gamma}\nonumber\\&+&
\mathcal{V}_5m_{N}^2(\gamma_\mu
C)_{\alpha\beta}(i\sigma^{\mu\nu}x_\nu\gamma_5N)_{\gamma} +
\mathcal{V}_6m_{N}^3(\rlap/xC)_{\alpha\beta}(\rlap/x\gamma_5N)_{\gamma}
+(\mathcal{A}_1\nonumber\\
&+&\frac{x^2m_{N}^2}{4}\mathcal{A}_1^M)(\rlap/p\gamma_5
C)_{\alpha\beta}N_{\gamma}+
\mathcal{A}_2m_{N}(\rlap/p\gamma_5C)_{\alpha\beta}(\rlap/xN)_{\gamma}+
\mathcal{A}_3m_{N}(\gamma_\mu\gamma_5 C)_{\alpha\beta}(\gamma^\mu
N)_{\gamma}\nonumber\\&+&
\mathcal{A}_4m_{N}^2(\rlap/x\gamma_5C)_{\alpha\beta}N_{\gamma}+
\mathcal{A}_5m_{N}^2(\gamma_\mu\gamma_5
C)_{\alpha\beta}(i\sigma^{\mu\nu}x_\nu N)_{\gamma}+
\mathcal{A}_6m_{N}^3(\rlap/x\gamma_5C)_{\alpha\beta}(\rlap/x
N)_{\gamma}\nonumber\\&+&(\mathcal{T}_1+\frac{x^2m_{N}^2}{4}\mathcal{T}_1^M)(p^\nu
i\sigma_{\mu\nu}C)_{\alpha\beta}(\gamma^\mu\gamma_5
N)_{\gamma}+\mathcal{T}_2m_{N}(x^\mu p^\nu
i\sigma_{\mu\nu}C)_{\alpha\beta}(\gamma_5 N)_{\gamma}\nonumber\\&+&
\mathcal{T}_3m_{N}(\sigma_{\mu\nu}C)_{\alpha\beta}(\sigma^{\mu\nu}\gamma_5
N)_{\gamma}+
\mathcal{T}_4m_{N}(p^\nu\sigma_{\mu\nu}C)_{\alpha\beta}(\sigma^{\mu\rho}x_\rho\gamma_5
N)_{\gamma}\nonumber\\&+& \mathcal{T}_5m_{N}^2(x^\nu
i\sigma_{\mu\nu}C)_{\alpha\beta}(\gamma^\mu\gamma_5 N)_{\gamma}+
\mathcal{T}_6m_{N}^2(x^\mu p^\nu
i\sigma_{\mu\nu}C)_{\alpha\beta}(\rlap/x\gamma_5
N)_{\gamma}\nonumber\\&+&
\mathcal{T}_7m_{N}^2(\sigma_{\mu\nu}C)_{\alpha\beta}(\sigma^{\mu\nu}\rlap/x\gamma_5
N)_{\gamma}+
\mathcal{T}_8m_{N}^3(x^\nu\sigma_{\mu\nu}C)_{\alpha\beta}(\sigma^{\mu\rho}x_\rho\gamma_5
N)_{\gamma} \, \, ,
\end{eqnarray}
where, the calligraphic functions, which are functions of the scalar
product $px$ and the parameters $a_i$, $i=1,2,3$, can be expressed
in terms of the nucleon distribution amplitudes (DA's) with the
increasing twist. The distribution amplitudes with different twist
are given explicitly in Tables \ref{tab:1}, \ref{tab:2},
\ref{tab:3}, \ref{tab:4} and \ref{tab:5}:
\begin{table}[h]
\centering
\begin{tabular}{|c|} \hline
$\mathcal{S}_1 = S_1$\\\cline{1-1}\hline
 $2px\mathcal{S}_2=S_1-S_2$ \\\cline{1-1}
   \end{tabular}
\vspace{0.3cm} \caption{Relations between the calligraphic functions
and proton scalar DA's.}\label{tab:1}
\end{table}
\begin{table}[h]
\centering
\begin{tabular}{|c|} \hline
  $\mathcal{P}_1=P_1$\\\cline{1-1}
  $2px\mathcal{P}_2=P_1-P_2$ \\\cline{1-1}
   \end{tabular}
\vspace{0.3cm} \caption{Relations between the calligraphic functions
and proton pseudo-scalar DA's.}\label{tab:2}
\end{table}
\begin{table}[h]
\centering
\begin{tabular}{|c|} \hline
  $\mathcal{V}_1=V_1$ \\\cline{1-1}
  $2px\mathcal{V}_2=V_1-V_2-V_3$ \\\cline{1-1}
  $2\mathcal{V}_3=V_3$ \\\cline{1-1}
  $4px\mathcal{V}_4=-2V_1+V_3+V_4+2V_5$ \\\cline{1-1}
  $4px\mathcal{V}_5=V_4-V_3$ \\\cline{1-1}
  $4(px)^2\mathcal{V}_6=-V_1+V_2+V_3+V_4
 + V_5-V_6$ \\\cline{1-1}
 \end{tabular}
\vspace{0.3cm} \caption{Relations between the calligraphic functions
and proton vector DA's.}\label{tab:3}
\end{table}
\begin{table}[h]
\centering
\begin{tabular}{|c|} \hline
  $\mathcal{A}_1=A_1$ \\\cline{1-1}
  $2px\mathcal{A}_2=-A_1+A_2-A_3$ \\\cline{1-1}
   $2\mathcal{A}_3=A_3$ \\\cline{1-1}
  $4px\mathcal{A}_4=-2A_1-A_3-A_4+2A_5$ \\\cline{1-1}
  $4px\mathcal{A}_5=A_3-A_4$ \\\cline{1-1}
  $4(px)^2\mathcal{A}_6=A_1-A_2+A_3+A_4-A_5+A_6$ \\\cline{1-1}
 \end{tabular}
\vspace{0.3cm} \caption{Relations between the calligraphic functions
and proton axial vector DA's.}\label{tab:4}
\end{table}
\begin{table}[h]
\centering
\begin{tabular}{|c|} \hline
  $\mathcal{T}_1=T_1$ \\\cline{1-1}
  $2px\mathcal{T}_2=T_1+T_2-2T_3$ \\\cline{1-1}
   $2\mathcal{T}_3=T_7$ \\\cline{1-1}
  $2px\mathcal{T}_4=T_1-T_2-2T_7$ \\\cline{1-1}
  $2px\mathcal{T}_5=-T_1+T_5+2T_8$ \\\cline{1-1}
  $4(px)^2\mathcal{T}_6=2T_2-2T_3-2T_4+2T_5+2T_7+2T_8$ \\\cline{1-1}
  $4px \mathcal{T}_7=T_7-T_8$\\\cline{1-1}
  $4(px)^2\mathcal{T}_8=-T_1+T_2 +T_5-T_6+2T_7+2T_8$\\\cline{1-1}
 \end{tabular}
\vspace{0.3cm} \caption{Relations between the calligraphic functions
and proton tensor DA's.}\label{tab:5}
\end{table}

  One can expresses the distribution amplitudes $F(a_ipx)$=  $S_i$,
$P_i$, $V_i$, $A_i$, $T_i$ as:
\begin{equation}\label{dependent1}
F(a_ipx)=\int dx_1dx_2dx_3\delta(x_1+x_2+x_3-1) e^{-ip
x\Sigma_ix_ia_i}F(x_i)\; .
\end{equation}
Here $x_{i}$ with $i=1,~2,~3$ corresponds to the longitudinal
momentum fractions carried by the quarks.

Using the expressions for the heavy quark propagator and nucleon
distribution amplitudes and performing integral over $x$ the expression for the
correlation function in QCD or theoretical side is obtained.
Equating the corresponding structures from both representations of
the correlation function and applying Borel  transformation with
respect to $(p+q)^2$ to suppress the contribution of the higher
states and continuum, one can obtain sum rules for the  form factors
$f_{1}$, $f_{2}$, $f_{3}$, $g_{1}$, $g_{2}$ and $g_{3}$. Finally, to subtract the contribution of the higher states and the continuum,
quark-hadron duality is assumed.

In heavy
quark effective theory (HQET), the heavy quark symmetry reduces the
number of independent form factors to two namely, $F_1$ and $F_2$
\cite{Mannel,alievozpineci}, i.e.,
\begin{eqnarray}\label{matrixel1111}
\langle N(p)\mid \bar u\Gamma b\mid \Sigma_b(p+q)\rangle&=&\bar
N(p)[F_1(Q^2)+\not\!vF_2(Q^2)]\Gamma u_{\Sigma_b}(p+q),\nonumber\\
\end{eqnarray}
where, $\Gamma$ is any Dirac structure and
$\not\!v=\frac{\not\!p+\not\!q}{m_{\Sigma_b}}$. Comparison between
Eq. (\ref{matrixel1111}) with the general definition of the form
factors in Eq. (\ref{matrixel1}) leads to the following relations
among the form factors in HQET limit \cite{Chen,ozpineci}
\begin{eqnarray}\label{matrixel22222}
g_1 = f_1=F_{1} + \frac{m_N}{m_{\Sigma_b}}F_{2}\nonumber\\
g_2 = f_2 = g_3 = f_3=\frac{F_2}{m_{\Sigma_b}}
\end{eqnarray}
Our calculations show that the deviation from the relations $g_1 =
f_1$ and  $g_2 = f_2 = g_3 = f_3$ are negligible in the case of HQET
limit. However, when we consider  finite mass, the violation is
$(10-20)^0/_0$ for $Q^2>0$ and turns out to be large for  $Q^2<0$
values. The explicit expressions for the form factors are very
lengthy, so considering the above relations, we will present only
the expressions for $f_1$ and $f_2$ in the Appendix--A. However, we
will give the extrapolation of all  form factors in finite mass in
terms of $Q^2$ in the numerical analysis section.

From the explicit expressions of the form factors, it is clear that
we need to know the expression for the residue of the $\Sigma_b$
baryon. The residue $\lambda_{\Sigma_b}$ is determined from sum rule
and its expression is given in \cite{Ozpineci1} as:
\begin{eqnarray}\label{residu2}
-\lambda_{\Sigma_{b}}^{2}e^{-m_{\Sigma_{b}}^{2}/M^{2}}&=&\int_{m_{b}^{2}}^{s_{0}}e^{\frac{-s}{M^{2}}}\rho(s)ds+e^{\frac{-m_b^2}{M^{2}}}\Gamma,
\end{eqnarray}
with
\begin{eqnarray}\label{residurho1}
\rho(s)&=&(<\overline{d}d>+<\overline{u}u>)\frac{(\beta^{2}-1)}{64
\pi^{2}}\Bigg\{\frac{m_{0}^{2}}{4 m_{b}}
(6\psi_{00}-13\psi_{02}-6\psi_{11})+3m_{b}(2\psi_{10}-\psi_{11}-\psi_{12}+2\psi_{21})\Bigg\}
\nonumber\\&+&\frac{ m_{b}^{4}}{2048 \pi^{4}}
[5+\beta(2+5\beta)][12\psi_{10}-6\psi_{20}+2\psi_{30}-4\psi_{41}+\psi_{42}
-12 ln(\frac{s}{m_{b}^{2}})],\nonumber\\
\end{eqnarray}

\begin{eqnarray}\label{lamgamma1}
\Gamma&=&\frac{
(\beta-1)^{2}}{24}<\overline{d}d><\overline{u}u>\left[\vphantom{\int_0^{x_2}}\right.\frac{m_{b}^{2}m_{0}^{2}}{2
M^{4}} +\frac{m_{0}^{2}}{4 M^{2}}-1\Bigg] ,
\end{eqnarray}
where, $s_0$ is continuum threshold, $M^2$ is the Borel mass
parameter and $\psi_{nm}=\frac{(s-m_b^2)^n}{s^m(m_b^2)^{n-m}}$ are
some dimensionless functions.
\section{Numerical results}
This section is devoted to the numerical analysis for the form
factors and  total decay rate  for $\Sigma_{b}\longrightarrow N
\ell\nu$ transition. Some input parameters used in the analysis of
the sum rules for the form factors are $\uu(1~GeV) = \dd(1~GeV)=
-(0.243)^3~GeV^3$, $m_N = 0.938~GeV$, $m_b = 4.7~GeV$,
$m_{\Sigma_{b}} = 5.805~GeV$, and $m_0^2(1~GeV) = (0.8\pm0.2)~GeV^2$
\cite{Belyaev}. The nucleon DA's are the main input parameters,
whose explicit expressions can be found in \cite{Lenz}. These DA's
contain 8 independent  parameters $f_{N},~\lambda_{1},
~\lambda_{2},~V_{1}^{d},~A_{1}^{u},~f_{1}^{d},~f_{1}^{u}$ and
$f_{2}^{d}$. These parameters have been calculated also in
\cite{Lenz} within the light cone QCD sum rules.  Recently, most of
these parameters have been calculated in the framework of the
lattice QCD \cite{Gockeler1,Gockeler2,QCDSF}. We will use these two
sets of data from        QCD sum rules and lattice QCD and for each
parameter which have not been calculated in lattice, we will use the
values from QCD sum rules prediction. These parameters are given in
Table \ref{kazem}.
\begin{table}[h]
\centering
\begin{tabular}{|c||c|c|} \hline
& QCD sum rules \cite{Lenz} & Lattice QCD
\cite{Gockeler1,Gockeler2,QCDSF}
\\\cline{1-3} \hline\hline
$f_{N}$ & $(5.0\pm0.5)\times10^{-3}~GeV^{2}$ &
$(3.234\pm0.063\pm0.086)\times10^{-3}~GeV^{2}$
\\\cline{1-3} $\lambda_{1}$ &$-(2.7\pm0.9)\times10^{-2}~GeV^{2}$ & $(-3.557\pm0.065\pm0.136)\times10^{-2}~GeV^{2}$ \\\cline{1-3}
 $\lambda_{2}$
&$(5.4\pm1.9)\times10^{-2}~GeV^{2}$&
$(7.002\pm0.128\pm0.268)\times10^{-2}~GeV^{2}$\\\cline{1-3}
$V_{1}^{d}$ &$0.23\pm0.03$& $0.3015\pm0.0032\pm0.0106$
\\\cline{1-3}
$A_{1}^{u}$ &$0.38\pm0.15$& $0.1013\pm0.0081\pm0.0298$\\\cline{1-3}
$f_{1}^{d}$ &$0.40\pm0.05$& $-$\\\cline{1-3} $f_{1}^{u}$
&$0.07\pm0.05$& $-$\\\cline{1-3} $f_{2}^{d}$ &$0.22\pm0.05$&
$-$\\\cline{1-3}
\end{tabular}
\vspace{0.8cm} \caption{The values of independent parameters
entering to the nucleon DA's. The first errors in lattice values are
statistical and the second errors represent the uncertainty due to
the chiral extrapolation and renormalization.} \label{kazem}
\end{table}

\begin{figure}[h!]
\begin{center}
\includegraphics[width=9cm]{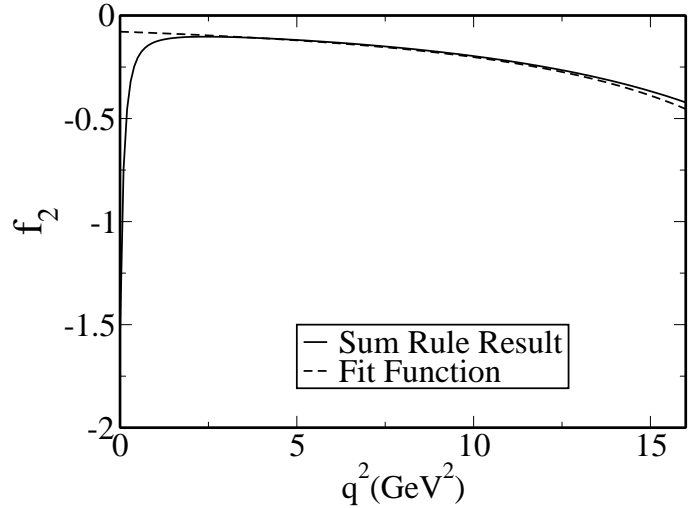}
\end{center}
\caption{The  dependency of $f_2$ (both actual sum rule result and fit parameterization) on $q^2$ for  QCD sum rules input parameters at $M^2=25~GeV^2$, $s_0=6.3^2~GeV^2$ and $\beta=5$.} \label{fig1}
\end{figure}

The sum rules for form factors also contain 3 auxiliary parameters
namely,  continuum threshold $s_0$, Borel mass parameter $M^2$ and
general parameter $\beta$ entering to the general current of the
$\Sigma_b$ baryon. These are not physical quantities, hence the form
factors should be independent of them.      Therefore, we look for
 working regions such that in these regions our results are practically independent of these mathematical objects.
 The continuum threshold, $s_0$ is not  completely arbitrary and it is related to the energy of the exited
states. Our numerical analysis for form factors show that the
results are weakly depend on $s_0$ in the interval,
$(m_{\Sigma_b}+0.5)^2\leq s_0\leq (m_{\Sigma_b}+0.7)^2$. In order to obtain the working region for  $\beta$, we plot the  form factors with respect to  $ cos\theta$ in the interval $-1\leq cos\theta\leq1$ which is corresponds to $-\infty\leq \beta\leq\infty$,
where $\beta=tan\theta$ and look for a region at which the dependency is weak. 
The common working
region for $\beta$ is obtained to be $-0.5\leq cos\theta\leq0.6$.  The Ioffe current which corresponds to
$cos\theta=-0.71$ is out of this region. The similar results have
been obtained in \cite{kazem3}. The lower limit on Borel mass
squared, $M^{2}$ is determined from condition that the contribution
of higher states and continuum to the correlation function should be
enough small, i.e., the contribution of the highest term
with power $1/M^2$ is less than, say, 20--25\% of the highest power of $M^{2}$. The  upper  limit of this parameter is acquired from the condition that
series of the light cone expansion with increasing twist should be
convergent. Generally, this means that the higher states, higher twists
and continuum contributions to the correlation function should be less than
40--50\% of the total value. Our numerical analysis show that both conditions are
satisfied in the region $15 ~GeV^{2}\leq M_{B}^{2}\leq 30~ GeV^{2}$,
which we will use in numerical analysis. Considering the above
requirements, we obtained that the form factors obey  the following
extrapolations in terms of $q^2$:
\begin{equation}\label{17au}
 f_{i}(q^2)[g_{i}(q^2)]=\frac{a}{(1-\frac{q^2}{m^2_{fit}})}+\frac{b}{(1-\frac{q^2}{m^2_{fit}})^2},
\end{equation}
 The values of the parameters
$a,~b$ and $m_{fit}$ are given in  Tables \ref{tab:7} and
\ref{tab:8} related to the  QCD sum rules and lattice QCD input
parameters, respectively. These parameterizations show that increasing in the value of  $q^2$ leads to increasing in the absolute value of the form factors and they have no pole inside the physical region. The values of $m_{fit}$ presents the pole outside the allowed region of $q^2$ and related to this and accordance to mesons, one can calculate the coupling constant  $g_{\Sigma_b\Sigma^*_bN}$, where, $\Sigma ^*_b$ can be considered as the exited state of $\Sigma_b$ baryon. For detailed analysis in this respect see \cite{damir1,damir2,damir3}. Note that, as we work near the light cone, $x^2\simeq0$, from the considered correlation function it is clear that our predictions at low $q^2$ are not reliable and we need the above parameterization to extend the results to full physical region. As an example, to show  how the actual sum rules results, and  the parameterization fit
to each other, we present the dependency of $f_2$  (both actual sum rule result and fit parameterization) on $q^2$ for QCD sum rules input parameters and at fixed values of auxiliary parameters in Fig. \ref{fig1}.

\begin{table}[h] \centering
\begin{tabular}{|c||c|c|c|} \hline
  & $a$ & $b$& $m_{fit}$\\\cline{1-4} \hline \hline
 $f_{1}$  & 0.13 & 0.005& 4.92\\\cline{1-4}
 $f_{2}$  &0.03 & -0.10& 5.40\\\cline{1-4}
  $f_{3}$  & -0.09 & -0.02& 4.92\\\cline{1-4}
 $g_{1}$ &0.20 & -0.05& 5.56\\\cline{1-4}
$g_{2}$ & -0.02 & 0.015& 5.96\\\cline{1-4} $g_{3}$ & -0.02 & -0.009&
5.65\\\cline{1-4}
  \end{tabular}
 \vspace{0.8cm}
\caption{Parameters appearing in the fit function for QCD sum rules
set of data.} \label{tab:7}
\end{table}
\begin{table}[h] \centering
\begin{tabular}{|c||c|c|c|} \hline
  & $a$ & $b$& $m_{fit}$\\\cline{1-4} \hline \hline
 $f_{1}$  & 0.19 & 0.004&4.88 \\\cline{1-4}
 $f_{2}$  & 0.038 &- 0.067&5.38 \\\cline{1-4}
  $f_{3}$  & -0.06 &-0.015 &4.93 \\\cline{1-4}
 $g_{1}$ & 0.25 & -0.064& 4.97\\\cline{1-4}
$g_{2}$ & -0.03 & -0.002& 5.97\\\cline{1-4} $g_{3}$ & -0.028 &-0.009
& 5.95\\\cline{1-4}
  \end{tabular}
 \vspace{0.8cm}
\caption{Parameters appearing in the fit function for  lattice QCD
set of data.} \label{tab:8}
\end{table}
The values of form factors at $q^2=0$ is also obtained as presented
in Table \ref{tab:melyas}.
\begin{table}[h]
\centering
\begin{tabular}{|c||c|c|} \hline
&For QCD sum rules input parameters & For lattice QCD input
parameters
\\\cline{1-3} \hline\hline
$f_{1}(0)$ & $0.14\pm0.05$ & $0.19\pm0.06$ \\\cline{1-3} $f_{2}(0)$
& $-0.08\pm0.03$ & $-0.029\pm0.010$ \\\cline{1-3} $f_{3}(0)$ &
$-0.11\pm0.04$ & $-0.076\pm0.028$\\\cline{1-3}$g_{1}(0)$
&$0.15\pm0.05$& $0.18\pm0.06$
\\\cline{1-3} $g_{2}(0)$ &$-0.036\pm0.012$&
$-0.033\pm0.011$\\\cline{1-3} $g_{3}(0)$ &$-0.032\pm0.011$&
$-0.037\pm0.012$\\\cline{1-3}
\end{tabular}
\vspace{0.8cm} \caption{The value of the form factors at $q^2=0$}.
\label{tab:melyas}
\end{table}
 Our next task is to calculate the total decay rate of $\Sigma_{b}\longrightarrow p \ell\nu$ transition in the whole
physical region, i.e., $ m_{l}^2 \leq q^2 \leq (m_{\Sigma_{b}} -
m_{N})^2$. The decay width  for such transition is given by the
following expression~\cite{Faessler,Pietschmann:1974ap}
\eq\label{Gamma_BiBf} \Gamma(\Sigma_b \to Pl \nu_l) =
\frac{G_F^2}{384 \pi^3 m_{\Sigma_b}^3} \ |V_{\rm bu}|^2 \,  \,
\int\limits_{m_l^2}^{\Delta^2} dq^2  \ (1 - m_l^2/q^2)^2 \
\sqrt{(\Sigma^2 - q^2) (\Delta^2 - q^2)} \ N(q^2) \en where \eq
N(q^2) &=& F_1^2(q^2) (\Delta^2 (4q^2 - m_l^2) + 2 \Sigma^2 \Delta^2
(1 + 2 m_l^2/q^2) - (\Sigma^2 + 2q^2) (2q^2 + m_l^2) )\nonumber\\[3mm]
&+& F_2^2(q^2) (\Delta^2 - q^2)(2 \Sigma^2 + q^2) (2q^2 +
m_l^2)/m_{\Sigma_b}^2
+ 3 F_3^2(q^2) m_l^2 (\Sigma^2 - q^2) q^2/m_{\Sigma_b}^2  \nonumber\\[3mm]
&+& 6 F_1(q^2) F_2(q^2) (\Delta^2 - q^2) (2 q^2 + m_l^2)
\Sigma/m_{\Sigma_{b}}
- 6 F_1(q^2) F_3(q^2)  m_l^2 (\Sigma^2 - q^2) \Delta/m_{\Sigma_b} \nonumber\\[3mm]
&+& G_1^2(q^2) (\Sigma^2 (4q^2 - m_l^2) + 2 \Sigma^2 \Delta^2
(1 + 2 m_l^2/q^2) - (\Delta^2 + 2q^2) (2q^2 + m_l^2) )\nonumber\\[3mm]
&+& G_2^2(q^2) (\Sigma^2 - q^2)(2 \Delta^2 + q^2) (2q^2 +
m_l^2)/m_{\Sigma_b}^2
+ 3 G_3^2(q^2) m_l^2 (\Delta^2 - q^2) q^2/m_{\Sigma_b}^2  \nonumber\\[3mm]
&-& 6 G_1(q^2) G_2(q^2) (\Sigma^2 - q^2) (2 q^2 + m_l^2)
\Delta/m_{\Sigma_b}
 +  6 G_1(q^2) G_3(q^2)  m_l^2 (\Delta^2 - q^2) \Sigma/m_{\Sigma_b}  \,.
\en
 Where $F_1(q^2)=f_1(q^2)$, $F_2(q^2)=m_{\Sigma_{b}}f_2(q^2)$, $F_3(q^2)=m_{\Sigma_{b}}f_3(q^2)$, $G_1(q^2)=g_1(q^2)$, $G_2(q^2)=m_{\Sigma_{b}}g_2(q^2)$,
 $G_3(q^2)=m_{\Sigma_{b}}g_3(q^2)$, $\Sigma  =
m_{\Sigma_{b}} + m_{p}$ and $\Delta  = m_{\Sigma_{b}} - m_{p}$. $G_F
= 1.17 \times 10^{-5}$ GeV$^{-2}$ is the Fermi coupling constant,
and $m_l$ is the leptonic (electron,  muon or tau) mass. For the
corresponding CKM matrix element $V_{ub}=(4.31\pm0.30)~10^{-3}$ is
used~\cite{Yao:2006px}. Our final results for total decay rates are
given in Table \ref{tab:9}.
\begin{table}[h]
\centering
\begin{tabular}{|c||c|c|c|c|} \hline
& For QCD sum rules input parameters &For lattice QCD input
parameters
\\\cline{1-3}\hline\hline
 $\Sigma_{b}\longrightarrow N e\nu_{e}$ & $(1.17\pm0.45) \times 10^{-15}$ & $(2.45\pm0.95) \times
10^{-15}$ \\\cline{1-3}  $\Sigma_{b}\longrightarrow N \mu\nu_{\mu}$
& $(1.19\pm0.45) \times 10^{-15}$ & $(2.46\pm0.95) \times
10^{-15}$\\\cline{1-3}$\Sigma_{b}\longrightarrow N \tau\nu_{\tau}$ &
$(7.23\pm2.65) \times 10^{-15}$ & $(4.82\pm1.75) \times
10^{-15}$\\\cline{1-3}
\end{tabular}
\vspace{0.8cm} \caption{Values of the total decay rate
 of the $\Sigma_{b}\longrightarrow N \ell\nu$}transition for  different leptons and two sets of input parameters obtained
 from QCD sum rules and lattice QCD. \label{tab:9}
\end{table}
From this Table, we see that the obtained results for the decay
rates are in the same order of magnitudes for two sets of input
parameters. The central values of the decay rate for $e$ and $\mu$
obtained using the lattice QCD input parameters are about 2 times
greater than that of the QCD sum rules input parameters while, for
$\tau$ case the result obtained by sum rules input parameters is about 1.5 time larger than the prediction acquired using the lattice input parameters. However, when we consider the
uncertainties, results obtained using both sets of input parameters
coincide for all leptons. Here, we should stress that as we mentioned before, the $\Lambda_{b}\rightarrow pl\bar\nu$ decay has  been studied in three point QCD sum rules  and HQET in \cite{yeni2} and using SU(3) symmetry and HQET in \cite{Datta}. Their predictions on the decay rate of the $\Lambda_{b}\rightarrow pl\bar\nu$ are, $1.35\times10^{-11}|V_{ub}|^2~GeV$ and $6.48\times10^{12}|V_{ub}|^2~s^{-1}$, respectively. In order  to have a sense of the order of  decay rates, we compare our average results presented in Table \ref{tab:9} with those predictions. Considering all results in the same unit,  we see that our average result is in the same order of magnitude with that of \cite{yeni2}, but one order of magnitude is greater than the \cite{Datta} prediction. For exact comparison the initial particles should be the same.

 In conclusion, using the most general form
of the interpolating currents of the heavy spin 1/2, $\Sigma_{b}$
baryon and distribution amplitudes
 of the nucleon, the transition form factors of the semileptonic $\Sigma_{b}\rightarrow Nl\nu$ were calculated in the framework of the light
 cone QCD sum rules. Ignoring the negligible  deviation, the form factors satisfied the HQET relations among the form factors.
 The obtained results for the related form factors were
 used to estimate the decay rate of this transition for two different sets of independent parameters  entering to expressions
 for the nucleon distribution amplitudes namely,  QCD sum rules and  lattice QCD input
 parameters. The obtained values for the decay rate for these two
 sets of data are approximately consistent with each other. Further improvements  would  be achieved by determining the next leading order
  QCD corrections to the nucleon distribution amplitudes.
\section{Acknowledgment}
The authors  thank T. M. Aliev for his useful discussions. This work
has been supported in part by the European Union (HadronPhysics2
project ``Study of strongly interacting matter''). K. A. thanks also TUBITAK, Turkish Scientific and Technical
Research Council, for their financial support provided under the
project 108T502.

\newpage

\section*{Appendix A}
In this section, we present the explicit expressions for the form
factors $f_1$ and $f_2$.
\begin{eqnarray}\label{f_{1}}
&&f_{1}(Q^{2})= \frac{1}{\sqrt{2}\lambda_{\Sigma_b}}
e^{m_{\Sigma_b}^{2}/M_{B}^{2}}\left(\vphantom{\int_0^{x_2}}\int_{t_{0}}^{1}dx_{2}\int_{0}^{1-x_{2}}dx_{1}
e^{-s(x_{2},Q^{2})/M_{B}^{2}}\frac{1}{2\sqrt{2}}\left[\vphantom{\int_0^{x_2}}m_b\left\{\vphantom{\int_0^{x_2}}(1+3\beta){\cal
H}_{19}(x_i)\right.\right.\right. -2(-1+\beta){\cal
H}_{17}(x_i)\nonumber\\&&\left.-(3+\beta){\cal
H}_{5}(x_i)\left.\vphantom{\int_0^{x_2}}\right\}-m_Nx_2\left\{\vphantom{\int_0^{x_2}}{\cal
H}_{1_2,-11,-13,19_2,-5,7}(x_i)+\beta{\cal
H}_{11,13,-17_2,-19_8,3_2,5,-7}(x_i)\vphantom{\int_0^{x_2}}\right\}\vphantom{\int_0^{x_2}}\right]\nonumber\\&&
+\int_{t_0}^1dx_2\int_0^{1-x_2}dx_1\int_{t_0}^{x_2}dt_1e^{-s(t_1,Q^2)/M_{B}^{2}
}\left[\vphantom{\int_0^{x_2}}-\frac{m_N^4
m_b}{M_B^4t_1^3\sqrt{2}}(-1+\beta)x_2{\cal
H}_{22}(x_i)\right.\nonumber\\&&\left.-\frac{m_N^2}{M_B^4t_1^2
2\sqrt{2}}\left\{\vphantom{\int_0^{x_2}}m_N^3x_2\Big[(-1+\beta){\cal
H}_{-10,16}(x_i)+2\beta{\cal
H}_{24}(x_i)\Big]+\Big[4m_N^3x_2+m_b\{Q^2+s(t_1,Q^2)\}(-1+\beta)x_2\right.\right.\nonumber\\&&\left.\left.-m_N^2m_b(-1+\beta)(2+3x_2)\Big]{\cal
H}_{22}(x_i)\vphantom{\int_0^{x_2}}\right\} +\frac{m_N^2}{M_B^4 t_1
2\sqrt{2}}
\left\{\vphantom{\int_0^{x_2}}m_Nx_2\Big[Q^2+s(t_1,Q^2){\cal
H}_{16,-22_4}(x_i)+(-1+\beta){\cal H}_{10}(x_i)
\right.\right.\nonumber\\&&\left.-\beta{\cal H}_{16,24_2}(x_i)\Big]
+m_b(-1+\beta)\Big[Q^2(1+3x_2)+s(t_1,Q^2)(1+x_2)\Big]{\cal
H}_{22}(x_i)+m_N^2m_b\Big[x_2(1+3\beta){\cal
H}_{16}(x_i)\right.\nonumber\\&& \left.+2(-1+\beta){\cal
H}_{24}(x_i) +(3+\beta){\cal H}_{10}(x_i)-(-1+\beta)(3+x_2){\cal
H}_{22}(x_i)\Big] -m_N^3\Big[(-1+\beta)(1+x_2){\cal
H}_{10,-16}(x_i)\right.\nonumber\\&&
-\left.\left.2\{\beta(1+x_2){\cal H}_{24}(x_i)+(2+4x_2){\cal
H}_{22}(x_i)\}\Big]\vphantom{\int_0^{x_2}}\right\}
+\frac{m_N^2}{M_B^42\sqrt{2}}\left\{\vphantom{\int_0^{x_2}}-3m_b
Q^2(-1+\beta){\cal
H}_{22}(x_i)\right.\right.\nonumber\\&&\left.\left.+m_N^2m_b\Big[(-1+\beta){\cal
H}_{22,-24_2}(x_i)-(1+3\beta){\cal H}_{16}(x_i)-(3+\beta){\cal
H}_{10}(x_i)\Big]+m_N^3\Big[(-8+2t_1-2x_2){\cal
H}_{22}(x_i)\right.\right.\nonumber\\&&\left.\left.+(-1+\beta){\cal
H}_{10,-16}(x_i)-2\beta{\cal
H}_{24}(x_i)\Big]+m_N\Big[Q^2(-1+\beta)(-1+t_1-x_2){\cal
H}_{10,-16}(x_i)\right.\right.\nonumber\\&&\left.\left.+Q^2(-6t_1+6x_2+4+2\beta){\cal
H}_{22}(x_i)+2Q^2\beta(1-t_1+x_2){\cal
H}_{24}(x_i)\Big]\vphantom{\int_0^{x_2}}\right\}+\frac{m_N^3}{M_B^2t_1^2
2\sqrt{2}}\left\{\vphantom{\int_0^{x_2}}{\cal
H}_{6,-18_3,20}(x_i)\right.\right.\nonumber\\&&\left.\left.+(-1+\beta){\cal
H}_{12}(x_i)-{\cal
H}_{6,-13,18}(x_i)\vphantom{\int_0^{x_2}}\right\}+\frac{m_N}{M_B^2t_1
4\sqrt{2}}\left\{\vphantom{\int_0^{x_2}}[Q^2+s(t_1,Q^2)]\Big[(3+25\beta){\cal
H}_{20}(x_i) +2(-1+\beta){\cal
H}_{-6,12}(x_i)\Big]\right.\right.\nonumber\\&&
-\left.\left.(5+\beta){\cal H}_{18}(x_i)-m_N^2\Big[2(-1+\beta){\cal
H}_{-6,12}(x_i) -(11+3\beta){\cal H}_{18}(x_i)+(5+67\beta){\cal
H}_{20}(x_i)\Big]\right.\right.\nonumber\\&&
\left.\left.+2x_2\Big[(-1+\beta){\cal H}_{-10,16}(x_i) +\beta{\cal
H}_{24}(x_i)\Big]-m_Nm_b\Big[{\cal
H}_{6_6,-8_3,-9_3,12_2,14,15,-20_4,21_4}(x_i) +4x_2(-1+\beta){\cal
H}_{22}(x_i)\right.\right.\nonumber\\&& \left.\left.+\beta {\cal
H}_{6_2,-8,-9,12_6,14_3,15_3,20_4,-21_4}(x_i)\Big]\vphantom{\int_0^{x_2}}\right\}
+\frac{m_N}{M_B^24\sqrt{2}}\left\{\vphantom{\int_0^{x_2}}Q^2\Big[{\cal
H}_{-6_2,+12_2,18_9,-20_3}(x_i)+\beta{\cal
H}_{6_2,-12_2,18_3,-20_{47}}(x_i)\Big]\right.\right.\nonumber\\&&
\left.\left.+4m_N(-1+\beta){\cal H}_{22}(x_i) +s(t_1,Q^2)\Big[{\cal
H}_{18_3,-20}(x_i)+\beta{\cal
H}_{18,-20_{21}}(x_i)\Big]\right.\right.\nonumber\\&&\left.\left.
+m_N^2\Big[\beta {\cal
H}_{4_4,8,-9,-10_2,14,-15,16_2,-18,20_{41},-21_4,23_{16},24_4}(x_i)
+{\cal H}_{-2_4,-8,9,10_2,-14,15,-16_2,-18_3,20,-23_4}(x_i)
\right.\right.\nonumber\\&&\left.\left.+8(t_1-x_2){\cal
H}_{22}(x_i)\Big]\vphantom{\int_0^{x_2}}\right\}+\frac{m_N}{t_14\sqrt{2}}\left\{\vphantom{\int_0^{x_2}}2(-1+\beta){\cal
H}_{-6,12}(x_i) +(1+5\beta){\cal H}_{20}(x_i)-(3+\beta){\cal
H}_{18}(x_i)\vphantom{\int_0^{x_2}}\right\}\right.\nonumber\\&&
+\left.\frac{m_N}{4\sqrt{2}}\left\{\vphantom{\int_0^{x_2}}(1+21\beta){\cal
H}_{20}(x_i) -(3+\beta){\cal
H}_{18}(x_i)\vphantom{\int_0^{x_2}}\right\}\vphantom{\int_0^{x_2}}\right]+\int_{t_{0}}^{1}dx_{2}\int_{0}^{1-x_{2}}dx_{1}e^{-s_{0}/M_{B}^{2}}
\left[\vphantom{\int_0^{x_2}}
\frac{m_N^4t_0^2}{(Q^2+m_N^2t_0^2)^3\sqrt{2}}(t_0-x_2)\left\{\vphantom{\int_0^{x_2}}
\right.\right.\nonumber\\&&\left.\left. \left.
\Big[m_N^2m_b(-1+\beta)(-2+t_0)(-1+t_0)
+2m_N^3t_0\{2+(-4+t_0)t_0\}-2m_Nt_0^2\{Q^2(-2+3t_0)+(-2+t_0)s(s_0,Q^2)\}
\right.\right.\right.\nonumber\\&&\left.\left.\left.-m_b(-1+\beta)t_0\{Q^2(-1+3t_0)+(-1+t_0)s(s_0,Q^2)\}\Big]{\cal
H}_{22}(x_i)+m_Nt_0\Big(\{m_N^2(-1+\beta)(-1+t_0)
\right.\right.\right.\nonumber\\&&-\left.\left.\left.m_N
m_bt_0(3+\beta)+(-1+\beta)t_0[Q^2(-1+t_0)-s(s_0,Q^2)]\}{\cal
H}_{10}(x_i)-\Big[m_N^2(-1+\beta)(-1+t_0)+m_Nm_bt_0(1+3\beta)\right.\right.\right.\nonumber\\&&
\left.\left.\left.+(-1+\beta)t_0\{Q^2(-1+t_0)-s(s_0,Q^2)\}\Big]{\cal
H}_{16}(x_i)+2\Big[-m_Nm_bt_0(-1+\beta) +m_N^2\beta(1-t_0)+\beta
t_0(Q^2(1-t_0)\right.\right.\right.\nonumber\\&&
\left.\left.\left.+s(s_0,Q^2))\Big]{\cal
H}_{24}(x_i)\Big)\right.\vphantom{\int_0^{x_2}}\right\}
+\frac{m_N^2}{(Q^2+m_N^2t_0^2)^22\sqrt{2}}(t_0-x_2)\left\{\vphantom{\int_0^{x_2}}\Big[m_N^2m_b(-1+\beta)(-2+t_0)(-1+t_0)
\right.\right.\nonumber\\&&\left.\left.+2m_N^3t_0\{2+(-4+t_0)t_0\}-2m_Nt_0^2\{Q^2(-2+3t_0)+(-2+t_0)s(s_0,Q^2)\}
-m_bt_0(-1+\beta)\{Q^2(-1+3t_0)
\right.\right.\nonumber\\&&\left.\left.+(-1+t_0)s(s_0,Q^2)\}\Big]{\cal
H}_{22}(x_i)+m_Nt_0\Big[\{m_N^2(-1+\beta)(-1+t_0)-m_Nm_bt_0(3+\beta)
+(-1+\beta)t_0[Q^2(-1+t_0)\right.\right.\nonumber\\&&\left.\left.-s(s_0,Q^2)]\}{\cal
H}_{10}(x_i)-\{m_N^2(-1+\beta)(-1+t_0)
+m_Nm_bt_0(1+3\beta)+(-1+\beta)t_0(Q^2(-1+t_0)-s(s_0,Q^2))\}{\cal
H}_{16}(x_i)
\right.\right.\nonumber\\&&\left.\left.+2\{-m_Nm_bt_0(-1+\beta)+m_N^2\beta(1-t_0)
+\beta t_0[Q^2(1-t_0)+s(s_0,Q^2)]\}{\cal
H}_{24}(x_i)\Big]\vphantom{\int_0^{x_2}}\right\} \right.\nonumber\
\end{eqnarray}

\begin{eqnarray}
&&+\frac{m_N}{(Q^2+m_N^2t_0^2)4\sqrt{2}M_B^2t_0}\left.\left\{\vphantom{\int_0^{x_2}}2m_N(t_0-x_2)
\Big[m_N^2m_b(-1+\beta)(-2+t_0)(-1+t_0)
+2m_N^3t_0(2+(-4+t_0)t_0)\right.\right.\nonumber\\&&\left.\left.+m_bt_0(-1+\beta)\{Q^2(1-3t_0)+2M_B^2t_0+(1-t_0)s(s_0,Q^2)\}
+2m_Nt_0^2\{Q^2(2-3t_0)\}+2M_B^2t_0\right.\right.\nonumber\\&&\left.\left.+(2-t_0)s(s_0,Q^2)\Big]{\cal
H}_{22}(x_i)+t_0\Big[m_N^2M_B^2{\cal H}_{6_2,-12_2,-18_6,20_2}(x_i)
+\beta{\cal H}_{-6_2,12_2,-18_2,20_26}(x_i)\Big]+2m_N^4(1-t_0){\cal
H}_{10,16}(x_i)
\right.\right.\nonumber\\&&\left.\left.+m_N^2M_B^2t_0{\cal
H}_{-6_2,12_2,18_{11},-20_5}(x_i) +m_Nm_bM_B^2t_0{\cal
H}_{-6_6,8_3,9_3,-12_2,-14,-15,20_4,-21_4}(x_i) +M_B^2Q^2t_0{\cal
H}_{6_2,-12_2,-18_5,20_3}(x_i)
\right.\right.\nonumber\\&&+\left.\left.m_N^4\beta t_0{\cal
H}_{-10_2,16_2,24_4}(x_i) +m_N^2M_B^2\beta t_0{\cal
H}_{6_2,-12_2,18_3,-20_{67}}(x_i) +m_Nm_bM_B^2\beta t_0{\cal
H}_{-6_2,8,9,-12_6,-14_3,-15_3,-20_4,21_4}(x_i)\right.\right.\nonumber\\&&\left.\left.
+M_B^2Q^2\beta t_0{\cal
H}_{-6_2,12_2,-18,20_{25}}(x_i)+2m_N^4t_0{\cal
H}_{-10,16}(x_i)+m_N^2m_bt_0^2{\cal H}_{-10_6,-16_2,24_4}(x_i)
\right.\right.\nonumber\\&&\left.\left.+m_N^2M_B^2t_0^2{\cal
H}_{-2_4,-8,9,10_2,-14,15,-16_2,-18_3,20,-23_4}(x_i)
+2m_N^2Q^2t_0^2{\cal H}_{10,-16}(x_i)+M_B^2Q^2t_0^2{\cal
H}_{-6_2,12_2,18_9,-20_3}(x_i)
\right.\right.\nonumber\\&&\left.\left.+m_N^4\beta t_0^2{\cal
H}_{10_2,-16_2,-24_4}(x_i)-m_N^3m_b\beta t_0^2{\cal
H}_{10_2,16_6,24_4}(x_i)+m_N^2M_B^2\beta t_0^2{\cal
H}_{4_4,8,-9,-10_2,14,-15,16_2,-18,20_{41},-21_4,23_{16},24_4}(x_i)\right.\right.\nonumber\\&&\left.\left.+m_N^2Q^2\beta
t_0^2{\cal H}_{-10_2,16_2,24_4}(x_i)+M_B^2Q^2\beta t_0^2{\cal
H}_{6_2,-12_2,18_3,-20_{47}}(x_i)-2Q^2t_0^3{\cal
H}_{10,-16}(x_i)+Q^2\beta t_0^3{\cal
H}_{10_2,-16_2,-24_4}(x_i)\right.\right.\nonumber\\&&\left.\left.+M_B^2t_0s(s_0,Q^2){\cal
H}_{6_2,-12_2,-18_5,20_3}(x_i)+M_B^2\beta t_0 s(s_0,Q^2){\cal
H}_{-6_2,12_2,-18,20_{25}}(x_i)+2m_N^2t_0s(s_0,Q^2){\cal
H}_{10,-16}(x_i)\right.\right.\nonumber\\&&\left.\left.+M_B^2t_0^2s(s_0,Q^2){\cal
H}_{18_3,-20}(x_i)+m_N^2\beta t_0^2s(s_0,Q^2){\cal
H}_{-10_2,16_2,24_4}(x_i)+M_B^2\beta t_0^2s(s_0,Q^2){\cal
H}_{18,-20_{21}}(x_i)\right.\right.\nonumber\\&&\left.\left.-2m_N^2x_2\Big[\{(-1+\beta)(-1+t_0)-m_Nm_bt_0(3+\beta)
+(-1+\beta)t_0[-M_B^2-Q^2(1-t_0)+s(s_0,Q^2)]\}{\cal
H}_{10}(x_i)\right.\right.\nonumber\\&&\left.\left.-\Big[m_N^2(-1+\beta)(-1+t_0)+m_Nm_bt_0(1+3\beta)+(-1+\beta)t_0\{-M_B^2-Q^2(1-t_0)
-s(s_0,Q^2)\}\Big]{\cal
H}_{16}(x_i)\right.\right.\nonumber\\&&\left.\left.+2\Big[-m_Nm_bt_0(-1+\beta)+m_N^2\beta(1-t_0)+\beta
t_0\{M_B^2+Q^2(1-t_0)+s(s_0,Q^2)\}\Big]{\cal
H}_{24}(x_i)\Big]\vphantom{\int_0^{x_2}}\right\}\vphantom{\int_0^{x_2}}\right),~~~~~~~~~~~~~~~(A.1)\nonumber\
\end{eqnarray}
  \begin{eqnarray}\label{f_{2}}
&&f_{2}(Q^{2})= \frac{1}{\sqrt{2}\lambda_{\Sigma_b}}
e^{m_{\Sigma_b}^{2}/M_{B}^{2}}\left(\vphantom{\int_0^{x_2}}\int_{t_{0}}^{1}dx_{2}\int_{0}^{1-x_{2}}dx_{1}
e^{-s(x_{2},Q^{2})/M_{B}^{2}}\frac{1}{2\sqrt{2}x_2}\left[\vphantom{\int_0^{x_2}}{\cal
H}_{11,-17_{2},5}(x_i) -\beta{\cal
H}_{11,-17_{12},5}(x_i)\vphantom{\int_0^{x_2}}\right]
\right.\nonumber\\&&\left.+\int_{t_0}^1dx_2\int_0^{1-x_2}dx_1\int_{t_0}^{x_2}dt_1e^{-s(t_1,Q^2)/M_{B}^{2}
}\left[\vphantom{\int_0^{x_2}}-\frac{m_N^4
}{M_B^4t_1^3\sqrt{2}}(3+\beta)x_2{\cal
H}_{22}(x_i)+\frac{m_N^2}{M_B^4t_1^2
2\sqrt{2}}\left\{\vphantom{\int_0^{x_2}}m_N m_b
x_2\Big[(1+3\beta){\cal
H}_{16}(x_i)\right.\right.\right.\nonumber\\&&
\left.\left.+2(-1+\beta){\cal H}_{24}(x_i)+(3+\beta){\cal
H}_{10}(x_i)\Big]+2\Big[-m_N m_b
x_2(-1+\beta)-\{Q^2+s(t_1,Q^2)\}(3+\beta)x_2+m_N^2(3+\beta\right.\right.\nonumber\\&&\left.\left.+(5+\beta)x_2\Big]{\cal
H}_{22}(x_i)\vphantom{\int_0^{x_2}}\right\} +\frac{m_N^2}{M_B^4 t_1
2\sqrt{2}} \left\{\vphantom{\int_0^{x_2}}m_N m_b
\Big[(1+3\beta){\cal H}_{16}(x_i)+2(-1+\beta){\cal
H}_{24}(x_i)+(3+\beta){\cal
H}_{10}(x_i)\Big]\right.\right.\nonumber\\&&\left.\left.+2\Big[m_N
m_b
(1-\beta)-s(t_1,Q^2)(3+\beta+x_2)+m_N^2(5+\beta+x_2)-Q^2(3+\beta+(4+\beta)x_2)\Big]{\cal
H}_{22}(x_i) \vphantom{\int_0^{x_2}}\right\}
\right.\nonumber\\&&+\left.\frac{m_N^2}{M_B^4\sqrt{2}}\left\{\vphantom{\int_0^{x_2}}\Big[m_N^2-(4+\beta)Q^2-s(t_1,Q^2)\Big]{\cal
H}_{22}(x_i)\vphantom{\int_0^{x_2}}\right\}+\frac{m_N}{M_B^2t_1^2
2\sqrt{2}}\left\{\vphantom{\int_0^{x_2}}-m_b\Big[{\cal
H}_{6_3,12,-18,-20}(x_i)+\beta{\cal
H}_{6,12_3,18,20}(x_i)\Big]\right.\right.\nonumber\\&&\left.\left.-2
(2+\beta)x_2{\cal
H}_{22}(x_i)\vphantom{\int_0^{x_2}}\right\}+\frac{m_N^2}{M_B^2t_1
4\sqrt{2}}\left\{\vphantom{\int_0^{x_2}}{\cal
H}_{-2_4,-8,9,15,-18_2,-20_2,22_8,-23_4}(x_i) +(-1+\beta){\cal
H}_{14}(x_i)\right.\right.\nonumber\\&& \left.\left.-\beta{\cal
H}_{-4_4,-8,9,15,18_2,-20_{22},21_4,-22_4,-23_{16},}(x_i)-8{\cal
H}_{22}(x_i) \vphantom{\int_0^{x_2}}\right\}
+\frac{m_N^2\sqrt{2}}{M_B^2}{\cal
H}_{22}(x_i)\vphantom{\int_0^{x_2}}\right]\nonumber\\
&&+\left.\int_{t_{0}}^{1}dx_{2}\int_{0}^{1-x_{2}}dx_{1}e^{-s_{0}/M_{B}^{2}}\left[\vphantom{\int_0^{x_2}}
\frac{m_N^4t_0^2}{(Q^2+m_N^2t_0^2)^3\sqrt{2}}(t_0-x_2)\left\{\vphantom{\int_0^{x_2}}
-m_N m_b t_0\Big[(1+3\beta){\cal H}_{16}(x_i)+2(-1+\beta){\cal
H}_{24}(x_i)\right.\right.\right.\nonumber\\&&\left.\left.\left.+(3+\beta){\cal
H}_{10}(x_i)+2(m_N
m_b(-1+\beta)t_0+m_N^2(3+\beta-(5+\beta)t_0+t_0^2)+t_0(Q^2(3+\beta)-(4+\beta)t_0)
\right.\right.\right.\nonumber\\
&&\left.\left.\left.+(3+\beta-t_0)s(s_0,Q^2)\Big]{\cal
H}_{22}(x_i)\vphantom{\int_0^{x_2}}\right\}
-\frac{m_N^2}{(Q^2+m_N^2t_0^2)^22\sqrt{2}}(t_0-x_2)\left\{\vphantom{\int_0^{x_2}}m_N
m_b t_0\Big[(1+3\beta){\cal H}_{16}(x_i)+2(-1+\beta){\cal
H}_{24}(x_i)\right.\right.\right.\nonumber\\&&\left.\left.+(3+\beta){\cal
H}_{10}(x_i)\Big]+2\Big[-m_N
m_b(-1+\beta)t_0+Q^2t_0\{-3-\beta+(4+\beta)t_0\}+m_N^2\{-3+\beta(-1+t_0)-(-5+t_0)t_0\}\right.\right.\nonumber\\&&
\left.\left.+t_0(-3-\beta+t_0)s(s_0,Q^2)\Big]{\cal
H}_{22}(x_i)\vphantom{\int_0^{x_2}}\right\}
+\frac{m_N}{(Q^2+m_N^2t_0^2)4\sqrt{2}M_B^2t_0}\left\{\vphantom{\int_0^{x_2}}4m_N(t_0-x_2)
\Big[m_N m_b(-1+\beta)t_0
+m_N^2\{3+\beta\right.\right.\nonumber\
\end{eqnarray}
  \begin{eqnarray}\label{f_{2}}
&&\left.-(5+\beta)t_0+t_0^2\}+t_0\{M_B^2(2+\beta+2t_0)+Q^2\{3+\beta-(4+\beta)t_0\}
+(3+\beta-t_0)s(s_0,Q^2)\}{\cal
H}_{22}(x_i)\right.\nonumber\\&&\left. -t_0\Big[m_b M_B^2\{{\cal
H}_{6_6,12_2,-20_2}(x_i)+\beta{\cal H}_{6_2,12_6,20_2}(x_i)\}+m_N^2
m_b t_0\{{\cal H}_{10_6,16_2,-24_4}(x_i)+\beta{\cal
H}_{10_2,16_6,24_4}(x_i)\}\right.\nonumber\\&&\left.+m_N
M_B^2t_0\{{\cal H}_{2_4,8,-9,14,-15,20_2,23_4}(x_i)+\beta{\cal
H}_{-4_4,-8,9,-14,15,-20_{22},21_4,-23_{16}}(x_i)\}+2M_B^2\{m_b(-1+\beta)\right.\nonumber\\&&\left.\left.\left.+m_N(1+\beta)t_0\}{\cal
H}_{18}(x_i)-2m_N^2m_b x_2\{(3+\beta){\cal
H}_{10}(x_i)+(1+3\beta){\cal H}_{16}(x_i)+2(-1+\beta){\cal
H}_{24}(x_i)\}\Big]
\vphantom{\int_0^{x_2}}\right\}\vphantom{\int_0^{x_2}}\right]\vphantom{\int_0^{x_2}}\right\},~~~~(A.2)\nonumber\
\end{eqnarray}

where
\begin{eqnarray}
{\cal H}(x_i) &=& {\cal H}(x_1,x_2,1-x_1-x_2),
\nonumber \\
s(y,Q^2)&=&(1-y)m_{N}^2+\frac{(1-y)}{y}Q^2+\frac{m_b^2}{y},~~~~~~~~~~~~~~~~~~~~~~~~~~~~~~~~~~~~~~~~~~~~~~~~~~~~~~~~~~~~~~~~~~~~~~~~~~~~~~~~~~~(A.3)\nonumber\
\end{eqnarray}
and $t_0=t_{0}(s_{0},Q^2)$ is the solution of the equation
$s(t_{0},Q^2)=s_{0}$, and is given as
\begin{eqnarray}
t_{0}(s_{0},Q^2)=\frac{m_N^2-Q^2-\sqrt{-4m_N^2(m_b^2-Q^2)+(-m_N^2+Q^2-s_0)^2}+s_0}{2m_N^2}.~~~~~~~~~~~~~~~~~~~~~~~~~~~~~~~~~~~~~~~~~~~~~~~~~~(A.4)\nonumber\
\end{eqnarray}

In the above equations, we have used  the short hand notations for
the functions ${\cal H}_{\pm i_a,\pm j_b, ...}=\pm a{\cal H}_{i}\pm
b{\cal H}_{j}...$, and  ${\cal H}_{i}$ are defined in terms of the
distribution amplitudes as follows:
\begin{eqnarray}
&&{\cal H}_{1}=S_{1}~~~~~~~~~~~~~~~~~~~~~~~~~~~~~~~~~~~~~~~~{\cal
H}_{2}=S_{1,-2}\nonumber\\&&{\cal
H}_{3}=P_{1}~~~~~~~~~~~~~~~~~~~~~~~~~~~~~~~~~~~~~~~~{\cal
H}_{4}=P_{1,-2}\nonumber\\&&{\cal
H}_{5}=V_{1}~~~~~~~~~~~~~~~~~~~~~~~~~~~~~~~~~~~~~~~~{\cal
H}_{6}=V_{1,-2,-3}\nonumber\\&&{\cal
H}_{7}=V_{3}~~~~~~~~~~~~~~~~~~~~~~~~~~~~~~~~~~~~~~~~{\cal
H}_{8}=-2V_{1,-5}+V_{3,4}\nonumber\\&&{\cal
H}_{9}=V_{4,-3}~~~~~~~~~~~~~~~~~~~~~~~~~~~~~~~~~~~~~{\cal
H}_{10}=-V_{1,-2,-3,-4,-5,6}\nonumber\\&&{\cal
H}_{11}=A_{1}~~~~~~~~~~~~~~~~~~~~~~~~~~~~~~~~~~~~~~~{\cal
H}_{12}=-A_{1,-2,3}\nonumber\\&&{\cal
H}_{13}=A_{3}~~~~~~~~~~~~~~~~~~~~~~~~~~~~~~~~~~~~~~~{\cal
H}_{14}=-2A_{1,-5}-A_{3,4}\nonumber\\&&{\cal
H}_{15}=A_{3,-4}~~~~~~~~~~~~~~~~~~~~~~~~~~~~~~~~~~~~{\cal
H}_{16}=A_{1,-2,3,4,-5,6}\nonumber\\&&{\cal
H}_{17}=T_{1}~~~~~~~~~~~~~~~~~~~~~~~~~~~~~~~~~~~~~~~~{\cal
H}_{18}=T_{1,2}-2T_{3}\nonumber\\&&{\cal
H}_{19}=T_{7}~~~~~~~~~~~~~~~~~~~~~~~~~~~~~~~~~~~~~~~~{\cal
H}_{20}=T_{1,-2}-2T_{7}\nonumber\\&&{\cal
H}_{21}=-T_{1,-5}+2T_{8}~~~~~~~~~~~~~~~~~~~~~~~~~~{\cal
H}_{22}=T_{2,-3,-4,5,7,8}\nonumber \\&&{\cal
H}_{23}=T_{7,-8}~~~~~~~~~~~~~~~~~~~~~~~~~~~~~~~~~~~~~{\cal
H}_{24}=-T_{1,-2,-5,6}+2T_{7,8},~~~~~~~~~~~~~~~~~~~~~~~~~~~~~~~~~~~~~~~~~~~~~~~~~(A.5)\nonumber
\
\end{eqnarray}
 where for any distribution amplitudes,  $X_{\pm i,\pm j, ...}=\pm X_{i}\pm X_{j}...$ are also used.


\begin{thebibliography}{99}
\bibitem{Mattson} M. Mattson et al., (SELEX Collaboration), Phys. Rev. Lett. 89,
112001 (2002).
 \bibitem{Ocherashvili} A. Ocherashvili et al., (SELEX Collaboration), Phys.
Lett. B 628, 18 (2005).
\bibitem{Acosta} D. Acosta et al., (CDF Collaboration), Phys.
Rev. Lett. 96, 202001 (2006).
 \bibitem{Chistov} R. Chistov et al., (Belle
Collaboration), Phys. Rev. Lett. 97, 162001 (2006).
\bibitem{Aubert1} B. Aubert et
al., (BABAR Collaboration), Phys. Rev. Lett. 97, 232001 (2006);
Phys. Rev. Lett. 99, 062001 (2007); Phys. Rev. D 77, 012002 (2008).
 \bibitem{Abazov1} V.
Abazov et al., (D0 Collaboration), Phys. Rev. Lett. 99, 052001
(2007); Phys. Rev. Lett. 101, 232002 (2008).
 \bibitem{Aaltonen1} T. Aaltonen et al., (CDF Collaboration), Phys. Rev. Lett.
99, 052002 (2007); Phys. Rev. Lett. 99, 202001 (2007).
\bibitem{Solovieva} E. Solovieva et al., (Belle Collaboration), Phys. Lett. B  672, 1 (2009).
\bibitem{kazem1} T. M. Aliev, K. Azizi, A. Ozpineci, Nucl. Phys. B 808, 137 (2009).
  \bibitem{Shuryak}  E. V. Shuryak, Nucl. Phys. B198, 83 (1982).
\bibitem{Kiselev1}  V. V. Kiselev,  A. I. Onishchenko, Nucl. Phys. B581, 432 (2000).
 \bibitem{Kiselev2} V. V. Kiselev, A. E. Kovalsky, Phys. Rev. D 64, 014002 (2001).
\bibitem{Bagan1} E. Bagan, M. Chabab, H. G. Dosch,  S. Narison, Phys. Lett. B 278, 367
(1992); Phys. Lett. B  287, 176 (1992).
  \bibitem{Bagan3} E. Bagan, M. Chabab,  S. Narison, Phys. Lett. B 306, 350 (1993).
\bibitem{Duraes} F. O. Duraes, M. Nielsen, Phys. Lett. B 658, 40 (2007).
 \bibitem{Wang1} Z. G. Wang, Eur. Phys. J. C 54, 231 (2008).
\bibitem{Zhang1} J. R. Zhang,  M. Q. Huang, Phys. Rev. D 77, 094002
(2008); Phys. Rev. D 78, 094007 (2008); Phys. Rev. D 78, 094015
(2008); Phys. Lett. B 674, 28 (2009); arXiv:0904.3391v1 [hep-ph].
\bibitem{Grozin} A. G. Grozin, d O. I. Yakovlev, Phys. Lett. B 285, 254 (1992); B 291, 441 (1992).
\bibitem{Groote} S. Groote, J. G. Korner,  O. I. Yakovlev, Phys. Rev. D 55, 3016 (1997).
 \bibitem{Dai} Y. B. Dai, C. S. Huang, C. Liu, C. D. Lu, Phys. Lett. B 371, 99 (1996).
 \bibitem{Lee} J. P. Lee, C. Liu, H. S. Song, Phys. Lett. B 476, 303 (2000).
 \bibitem{Huang} C. S. Huang, A. L. Zhang, S. L. Zhu, Phys. Lett. B 492, 288 (2000).
\bibitem{Liu} X. Liu, H. X. Chen, Y. R. Liu, A. Hosaka, and S. L. Zhu, Phys. Rev. D 77, 014031 (2008).
\bibitem{Wang2} D. W. Wang, M. Q. Huang,  C. Z. Li, Phys. Rev. D 65, 094036
(2002); Phys. Rev. D 67, 074025 (2003); Phys. Rev. D 68, 034019
(2003).
 \bibitem{Ebert1} D. Ebert, R. N. Faustov, V. O. Galkin, Phys. Rev. D 72, 034026
 (2005); Phys. Lett. B 659, 612 (2008);
\bibitem{Ebert3} D. Ebert, R. N. Faustov, V. O. Galkin,  A. P. Martynenko, Phys. Rev. D 66, 014008 (2002).
\bibitem{Capstick} S. Capstick,  N. Isgur, Phys. Rev. D 34, 2809 (1986).
 \bibitem{Matrasulov} D. U. Matrasulov, M. M. Musakhanov,  T. Mori, Phys. Rev. C 61, 045204 (2000).
\bibitem{Gershtein} S. S. Gershtein, V. V. Kiselev, A. K. Likhoded, A. I. Onishchenko, Phys Rev. D 62, 054021 (2000).
 \bibitem{Kiselev3} V. V. Kiselev, A. K. Likhoded, O. N. Pakhomova,  V. A. Saleev, Phys. Rev. D  66, 034030 (2002).
 \bibitem{Vijande} J. Vijande, H. Garcilazo, A. Valcarce,  F. Fernandez, Phys. Rev. D 70, 054022 (2004).
\bibitem{Martynenko} A. P. Martynenko, Phys. Lett. B 663, 317 (2008).
\bibitem{Hasenfratz} P. Hasenfratz, R. R. Horgan, J. Kuti,  J. M. Richard, Phys. Lett. B 94, 401 (1980).
\bibitem{kazem2} T. M. Aliev, K. Azizi, A. Ozpineci, Phys. Rev. D 77, 114006 (2008).
\bibitem{kazem3} T. M. Aliev, K. Azizi, A. Ozpineci, Phys. Rev. D 79, 056005 (2009).
\bibitem{yeni1} R. S. Marques de Carvalho, F. S. Navarra1, M. Nielsen, E. Ferreira,  H. G. Dosch, Phys. Rev. D 60, 034009 (1999).
\bibitem{yeni2} C. -S. Huang, C. -F.  Qiao, H. -G. Yan, Phys. Lett. B 437 (1998) 403.

\bibitem{Datta}  A. Datta,  arXiv:hep-ph/9504429.
 \bibitem{Roberts} W. Roberts, M. Pervin, arXiv:0803.3350 [nucl-th].
\bibitem{Guo} X. H. Guo, K. W. Wei, X. H.  Wu, Phys. Rev. D 77, 036003 (2008).
 \bibitem{Cheng} H. Y. Cheng, C. K. Chua, Phys. Rev. D 75, 014006 (2007).
\bibitem{D.Ebert} D. Ebert, R. N. Faustov, V. O. Galkin, Phys. Rev. D 73 (2006) 094002.
\bibitem{albertus} C. Albertus, E. Hernandez, J. Nieves, Phys. Rev. D 71 (2005) 014012.
\bibitem{pervin} M.  Pervin, W. Roberts, S. Capstick, Phys. Rev. C 72 (2005) 035201; 74, 025205 (2006).

\bibitem{Lenz}  V. M.  Braun, A. Lenz, M. Wittmann, Phys. Rev. D  73 (2006) 094019.
\bibitem{Gockeler1}  M. Gockeler et al.,  QCDSF Collaboration, PoS LAT2007 (2007) 147, arXiv:0710.2489 [hep-lat].
\bibitem{Gockeler2}  M. Gockeler et al., Phys. Rev. Lett. 101 (2008) 112002,    arXiv:0804.1877 [hep-lat].
\bibitem{QCDSF}  V. M. Braun et al., QCDSF Collaboration, arXiv:0811.2712 [hep-lat].
\bibitem{Balitsky} I. I.  Balitsky,  V. M.  Braun,  Nucl.
Phys.     B311 (1989) 541.


\bibitem{17}  V. M.  Braun, A. Lenz, N. Mahnke, E. Stein, Phys.
Rev. D  65 (2002) 074011.
\bibitem{18}  V. M.  Braun, A. Lenz, M. Wittmann, Phys.
Rev. D  73 (2006) 094019;
  A.~Lenz, M.~Wittmann and E.~Stein,
        Phys.\ Lett.\  B  581 (2004) 199.
\bibitem{Braun1b}  V. Braun, R.~J.~Fries, N.~Mahnke and E.~Stein, Nucl. Phys.
B  589 (2000) 381.
\bibitem{8}  V. M.  Braun, A. Lenz, G. Peters, A. V. Radyushkin, 
Phys. Rev. D 73(2006)034020.
 \bibitem{Mannel} T. Mannel, W. Roberts and Z. Ryzak,
Nucl. Phys. B355 (1991) 38.

\bibitem{alievozpineci}  T. M. Aliev, A. Ozpineci, M. Savci, Phys. Rev. D 65 (2002)
115002.
\bibitem{Chen} C. H. Chen,  C. Q. Geng, Phys. Rev. D 63 (2001) 054005; Phys.
Rev. D 63 (2001) 114024; Phys. Rev. D 64 (2001) 074001.
 \bibitem{ozpineci}  T. M. Aliev, A. Ozpineci, M. Savci, C. Yuce , Phys. Lett. B 542 (2002)
 229.
\bibitem{Ozpineci1} K. Azizi, M. Bayar, A. Ozpineci, Phys. Rev. D 79, 056002
(2009).

\bibitem{Belyaev} V. M. Belyaev,  B. L.  Ioffe, JETP  56 (1982) 493.
\bibitem{damir1} D. Becirevic, A. B. Kaidalov, Phys. Lett. B 478, 417 (2000).
 \bibitem{damir2}P. Ball, R. Zwicky, Phys. Rev. D71 (2005) 014015.
\bibitem{damir3}  V. M. Belyaev, V. M. Braun, A. Khodjamirian, R. Ruckl, Phys. Rev. D 51, 6177
 (1995).

\bibitem{Faessler} A. Faessler, T. Gutsche, B. R. Holstein, M. A. Ivanov, J. G. Korner, V. E. Lyubovitskij, Phys. Rev. D 78, 094005 (2008).
\bibitem{Pietschmann:1974ap}
  H.~Pietschmann,
  Acta Phys.\ Austriaca Suppl.\   12, 1 (1974).
\bibitem{Yao:2006px} C. Amsler et al. [Particle Data Group], Phys. Lett. B 667 1 (2008).
\end{thebibliography}
\end{document}